\documentclass{aastex63}
\usepackage{colordvi}
 \usepackage{color}
 \newcommand{\B}[1]{{\color{blue}{#1}}}
 \newcommand{\R}[1]{{\color{red}{#1}}}

\begin{document}

\title{Heating and Eruption of a Solar Circular Ribbon Flare}

\correspondingauthor{Jeongwoo Lee}
\email{leej@njit.edu}

\author[0000-0002-5865-7924]{Jeongwoo Lee}
\affil{Institute for Space Weather Sciences, New Jersey Institute of Technology, University Heights, Newark, NJ 07102-1982, USA}
\affil{Center for Solar-Terrestrial Research, New Jersey Institute of Technology, University Heights, Newark, NJ 07102-1982, USA}
\affil{Big Bear Solar Observatory, New Jersey Institute of Technology, 40386 North Shore Lane, Big Bear City, CA 92314-9672, USA}

\author[0000-0002-6975-5642]{Judith T. Karpen}
\affil{Code 674, NASA Goddard Space Flight Center, Greenbelt, MD, 20771, USA}

\author[0000-0002-6178-7471]{Chang Liu}
\affil{Institute for Space Weather Sciences, New Jersey Institute of Technology, University Heights, Newark, NJ 07102-1982, USA}
\affil{Center for Solar-Terrestrial Research, New Jersey Institute of Technology, University Heights, Newark, NJ 07102-1982, USA}
\affil{Big Bear Solar Observatory, New Jersey Institute of Technology, 40386 North Shore Lane, Big Bear City, CA 92314-9672, USA}

\author[0000-0002-5233-565X]{Haimin Wang}
\affil{Institute for Space Weather Sciences, New Jersey Institute of Technology, University Heights, Newark, NJ 07102-1982, USA}
\affil{Center for Solar-Terrestrial Research, New Jersey Institute of Technology, University Heights, Newark, NJ 07102-1982, USA}
\affil{Big Bear Solar Observatory, New Jersey Institute of Technology, 40386 North Shore Lane, Big Bear City, CA 92314-9672, USA}

\begin{abstract}
We studied a circular-ribbon flare, SOL2014-12-17T04:51, with emphasis on its thermal evolution as determined by the Differential Emission Measure (DEM) inversion analysis of the extreme ultraviolet (EUV) images of the Atmospheric Imaging Assembly (AIA) instrument onboard the Solar Dynamics Observatory (SDO). Both temperature and emission measure start to rise much earlier than the flare, along with an eruption and formation of a hot halo over the fan structure. In the main flare phase, another set of ribbons forms inside the circular ribbon, and expands as expected for ribbons at the footpoints of a postflare arcade.  An additional heating event further extends the decay phase, which is also characteristic of some eruptive flares. The basic magnetic configuration appears to be a fan-spine topology, rooted in a minority-polarity patch surrounded by majority-polarity flux. We suggest that reconnection at the null point begins well before the impulsive phase, when the null is distorted into a breakout current sheet, and that both flare and breakout reconnection are necessary in order to explain the subsequent local thermal evolution and the eruptive activities in this confined magnetic structure.
Using local DEMs, we found a postflare temperature increase inside the fan surface,  indicating that the so-called EUV late phase is due to continued heating in the flare loops. 

\end{abstract}

\section{Introduction}

Circular ribbon flares were introduced to the solar community by comprehensive studies based on 1600 {\AA} UV continuum images from Transition Region And Coronal Explorer (TRACE), model-based interpretation, H$\alpha$ blue-wing images retrieved from the old films of Big Bear Solar Observatory (BBSO), and hard X-ray observations with the Reuven Ramaty High Energy Solar Spectroscopic Imager (RHESSI)  (Masson et al. 2009; Reid et al. 2012; Wang \& Liu 2012). Later studies used high-resolution EUV images from the Atmospheric Imaging Assembly (AIA) instrument onboard the Solar Dynamics Observatory (SDO) to detect changes in the corona during circular ribbon flares (Sun et al. 2013; Mandrini et al. 2014; Wang et al. 2014; Yang et al. 2015; Masson et al. 2017; Xu et al. 2017). Magnetograms show that circular ribbon flares occur in a common magnetic configuration on the Sun: an embedded bipole, also known as a fan-spine topology, in which a central parasitic magnetic field is encompassed by the opposite polarity, with a magnetic null point on top of a dome-shaped separatrix sitting on a quasi-circular ribbon (Antiochos 1990, Lau \& Finn 1990, Schrijver \& Title 2002, T\"or\"ok et al. 2009, Pontin et al. 2013). Observations have thus far shown that circular ribbon flares typically involve three ribbons: a circular ribbon, an inner ribbon, and a remote ribbon in the far end of the spine and surrounding field lines. The 3D structure offered by circular ribbon flares differs significantly from that of two-ribbon flares, which can be approximated by a 2D X-line configuration (e.g., Kopp \& Pneuman 1976). Accordingly, many 3D MHD modeling studies have investigated the phenomena of circular ribbon flares (e.g., Rickard \& Titov 1996; Galsgaard \& Nordlund 1997; Galsgaard et al. 2003;  Pontin \& Galsgaard 2007; Pontin et al. 2007; Masson et al. 2009).  Many coronal jets originate in similar, albeit smaller and simpler, magnetic configurations, so observational and numerical studies of jets can yield insight into key physical processes involved in eruptive circular ribbon flares (e.g., Pariat et al. 2009, 2010, 2015, 2016; Sterling et al. 2015; Karpen et al. 2017; Wyper \& DeVore 2016, Wyper et al. 2018).   

The magnetic breakout model (Antiochos et al. 1999; MacNeice et al. 2004; Lynch et al. 2008; DeVore \& Antiochos 2008; Karpen et al. 2012; Dahlin et al. 2019) has been proposed as a comprehensive explanation for solar eruptive events in multipolar systems containing filament channels (Wyper et al. 2017). Numerous observations exhibit features consistent with this universal model for jets (Sterling et al. 2015; Kumar et al. 2018, 2019; Moore et al. 2018; Panesar et al. 2018; Li et al. 2019) and CMEs/eruptive flares (Aulanier et al. 2000; Sterling \& Moore 2001a,b, 2004a,b; Manoharan \& Kundu 2003; Williams et al. 2005; Alexander et al. 2006; Mandrini et al. 2006; Sui et al. 2006; Joshi et al. 2007; Lin et al. 2010; Aurass et al. 2011). The vast majority of jets and CMEs/eruptive flares originate in such systems, including circular ribbon flares, so the breakout paradigm provides a well-understood and tested context for interpreting circular ribbon flares. The presence of a filament channel (FC) around the polarity inversion line (PIL), regardless of whether or not it supports a filament, signifies a source of magnetic free energy that can power an eruption.  In contrast, the fan field overlying the FC acts to hold down the energized FC field, enabling increased energy storage. The breakout model invokes magnetic reconnection in two key locations to disrupt this force balance: {\it breakout reconnection} at a current sheet formed at a stressed null, and {\it flare reconnection} in the current sheet formed behind the rising sheared core field. The sequence of activity associated with a breakout event, as established by the above observations and simulations, begins with the distortion of the null at the fan-spine junction into a 3D current patch, which occurs easily when the flux system is stressed by the ambient photospheric shuffling. As this breakout current sheet evolves, slow reconnection through the breakout current sheet steadily converts closed fan flux to open (or effectively open) flux outside the separatrix, removing restraints on the sheared filament-channel field and enabling its slow rise. As the oppositely directed flux beneath the rising core is drawn together, the flare current sheet (FCS) forms and eventually starts to reconnect, creating a growing flux rope inside the fan.  During this rise phase, weak jets are emitted from the breakout current sheet and the FCS transitions from slow to fast reconnection, but explosive activity only begins when the upper edge of the flux rope encounters the breakout current sheet, initiating fast reconnection there.  Depending on the ratio of overlying flux to axial (sheared) flux, either a jet or a CME (or a combination ) is ejected (Kumar et al. 2020, in preparation). For circular ribbon flares the presence of both flare and circular ribbons clearly indicates that both types of reconnection are at work.

In this paper, we study the SOL2014-12-17T04:51 flare, which appears to possess a circular ribbon. The magnetic context of this event has been presented by Liu et al. (2019), who performed a nonlinear force-free field extrapolation from the HMI magnetograms.  They found evidence for several flux ropes around the fan field, and interpreted the flare activity and the eruption based on the interactions between the flux ropes and the embedded bipole.  In the present study, we focus on the spatio-temporal evolution of temperature obtained from the differential emission measure (DEM) inversion analysis of the AIA/SDO data (Cheung et al. 2015).  Tracking the temperature in each element of the circular ribbon flare is useful in understanding energy release and transfer, with implications for the magnetic eruption mechanism. It will also help us detecting thermal activities in EUV-dimming regions, which has received little attention in prior studies on circular ribbon flares. Although the thermal structure does not directly reveal the underlying magnetic structure, it traces the energy release as a consequence of magnetic reconnection, and therefore complements the abovementioned magnetic-field--based studies. Another interesting aspect of this event is the extended late-phase emission, which involves a second emission enhancement in warm coronal extreme-ultraviolet (EUV) lines. The mechanism for the late-phase EUV activity is considered to be either additional heating (Woods et al. 2011, Dai et al. 2013, Hock et al. 2019) or the long--cooling timescale (Liu et al. 2013). Specific proposed mechanisms include a two-loop system closely linked by the fan-spine topology (Sun et al. 2013), or continued reconnection behind the CME of stretched, unsheared flux  (Woods et al. 2011, Hock et al. 2019).

The plan of this paper is as follows: we present the (E)UV observations in \S 2, and the DEM inversion results in \S 3.  \S 4 describes our interpretation of the results.

\begin{figure}[tbh]  
\plotone{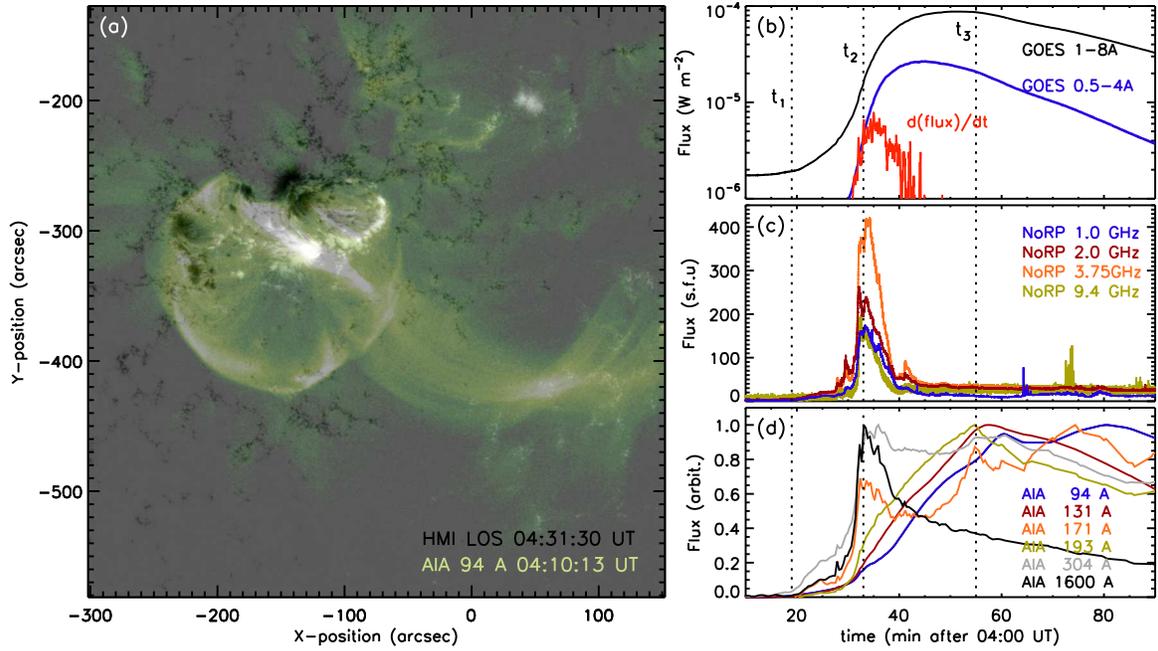} 
\caption{
NOAA AR 2242 and lightcurves of the SOL2014-12-17T04:51 flare.
Left panel: HMI magnetogram overlaid with AIA 94 {\AA} intensity.
Right panels: GOES soft X-ray lightcurves alog with its time derivative (top), NoRP microwave flux time profiles (middle), and normalized (E)UV fluxes from the FOV set by the left panel image.
The three vertical guide lines represent the time of the circular ribbon activation ($t_1$), the flare maximum ($t_2$), and the time of the maximum EUV line fluxes ($t_3$), respectively. }
\end{figure}

\section{Observation}

We study the SOL2014-12-17T04:51 flare that occurred in NOAA active region (AR) 12242 located close to the disk center (S20$^{\rm o}$, E9$^{\rm o}$), because the quasi-circular ribbons as well as the dome-like fan structure and outer spine in this event could be identified without ambiguity in the SDO/AIA images. 

\subsection{Data Overview}

We present an overview of the active region and flare lightcurves in Figure 1. Figure 1a shows the AIA 94 {\AA} image in the early phase overlaid over the HMI magnetogram, which shows a central positive-polarity region surrounded by the negative-polarity fields, forming a closed quasi-circular major polarity inversion line (PIL). The negative fields are then connected to a remote positive-polarity region in the west across a secondary PIL. The event therefore possesses major magnetic elements of a typical circular ribbon flare (e.g., Masson et al. 2009, Wang \& Liu 2012). According to the GOES 1--8 {\AA} lightcurve (Fig. 1b), this flare started at 04:25 UT, peaked at 04:51 UT at GOES class M8.7, and ended at 05:20 UT (defined as the time when the GOES flux decays to 50\% of its peak value) after a long duration.  The Nobeyama Radio Polarimeters (NoRP) detected microwave bursts during this period, with the largest starting around 04:20 UT and reaching maximum at 04:33 UT   (Fig. 1c). The Reuven Ramaty High Energy Solar Spectroscopic Imager (RHESSI) missed most of this flare, so hard X-ray data are not included in this study. In the AIA EUV lightcurves (Fig. 1d), the 304 {\AA} and 1600 {\AA} fluxes show an early activation at around 04:20 UT and peak at 04:33 UT together with the microwave bursts. Therefore, the 1600 {\AA} flux evolves like the microwave bursts. The 304 {\AA} intensity rises as impulsively as the 1600 {\AA} flux, but decays as slowly as other EUV lines. The EUV fluxes rise together after the impulsive phase and continue to remain high through the extended decay phase, which is longer than an hour, consistent with the soft X-ray burst. The Large Angle and Spectrometric Coronagraph (LASCO; Brueckner et al. 1995) on board the Solar and Heliospheric Observatory detected a halo CME associated with the flare, with the fastest moving segment of the CME leading edge at a linear speed of 587 km s$^{-1}$ located at the position angle of 162$^{\rm o}$  (the CME catalog: Yashiro et al. 2004). The relationship of the CME to this flare has been studied by Liu et al. (2019) using a magnetic field analysis and modeling. In this study, we analyze the SDO data from 04:10 UT to 05:30 UT in order to investigate the preflare brightenings, impulsive flaring, eruption, and late-phase extended emission in that period.

\begin{figure}[tbh]  
\plotone{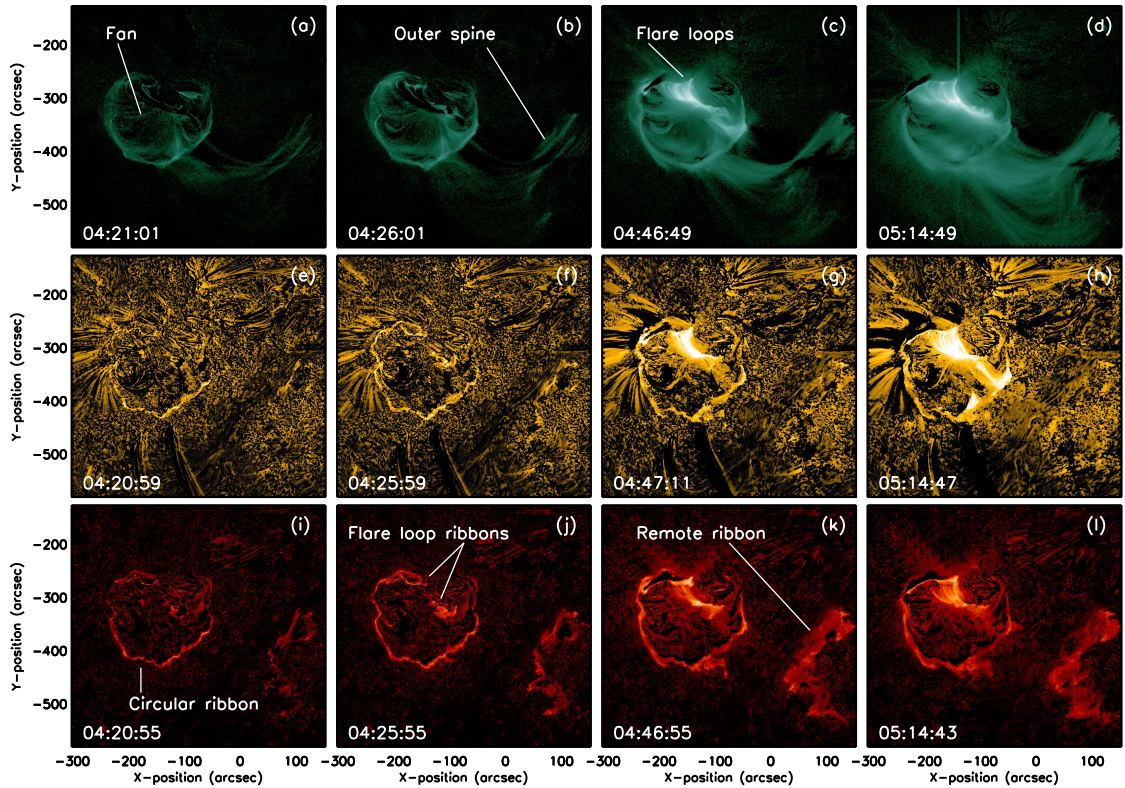} 
\caption{
SDO/AIA (E)UV base-difference images in three channels, 94 {\AA}, 171 {\AA}, and 304 {\AA} from top to bottom, at four selected times: from left to right, the preflare phase, the maximum phase, the secondary maximum, and the later postflare phase. A single preflare image (around 04:10 UT) is subtracted from all images in each channel. Major topological components of the circular ribbon flare are denoted. The accompanying animation shows temporal evolution of AIA images in six channels including 131 {\AA}, 335 {\AA}, and 193 {\AA} in addition to the above.}
\end{figure}

\subsection{EUV Emissions}

Figure 2 shows selected AIA base-difference images in three wavelengths: 94, 171, and 304 {\AA} from top to bottom. From left to right, the time corresponds to the preflare phase, the maximum phase, the secondary maximum phase, and the postflare phase. In each channel, a single preflare image (around 04:10 UT) is subtracted from all subsequent images for better visibility of the relevant features.  The preflare 94 {\AA} images (Fig. 2a--d) show a relatively uniform hemispheric structure with a spine-like structure attached.  

After 04:30 UT the 94 {\AA} image shows increased emission over the northern part of the AR, which expands like a postflare arcade. The outer spine halo is also well visible in the 94 {\AA} images, indicating that hot plasma is lingering in and around the outer spine. Later images seem to show braided field lines in the outer spine. These features look much thicker than expected for the fan surface and spines, which are infinitely thin structures by definition, so we call them haloes. It is noticeable that the fan halo in the postflare phase is much more developed than in the preflare phase. 

Similar morphology is found in the 131 {\AA} and 193 {\AA} images, but more complex morphology is  seen in  the 171 {\AA}, 193 {\AA}, and 211 {\AA} images. The EUV  images at 131 {\AA}, 335 {\AA}, and 193 {\AA} are not shown in Figure 2, but included in the animation accompanying Figure 2. In particular, the 171 {\AA} images do not yield the impression of the fan-spine structure (Fig. 2e--h). It may be that they show both the ribbon structure as in the 304 {\AA} images and the fan structure as in the 94 {\AA} images together. Although all features in the 94 {\AA} images may not belong to the fan surface, the hemispheric structure tends to be more clearly visible in the higher temperature lines and less obvious in the low temperature lines. It is thus likely that the fan is filled with hot plasma. 

The circular ribbon and remote footpoints are best visible in the 304 {\AA} images (Fig. 2i--l). The circular ribbon instantly turned on at $\sim$04:20 UT and stayed steady until the next enhancement at flare onset $\sim$04:32 UT. The southern section of the ribbon most clearly shows the quasi-circular shape, while the northeastern section is more complex and bulky. The northwestern section of the circular ribbon was very faint and remained so even during the flare. Therefore the whole ribbon is not evenly visible, which is not uncommon. This event does not exhibit sequential EUV brightening along the circular ribbon, as found in some other circular ribbon flares (Masson et al. 2009, Reid et al. 2012). 

\begin{figure}[tbh]  
\plotone{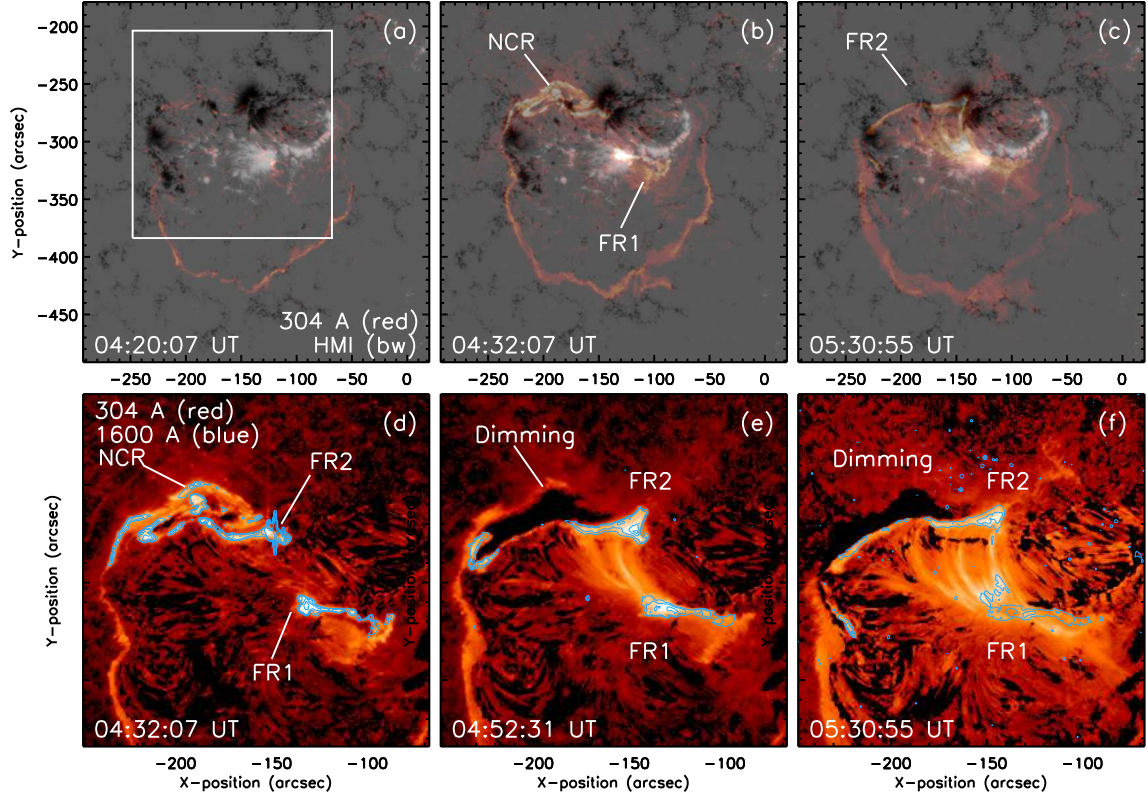}
\caption{Circular ribbon (NCR) and flare ribbons (FR1 and FR2). Top row: the 304 {\AA} base-difference images are plotted (red color table) over the HMI longitudinal magnetograms (black--white) to show the ribbons at three times. The white box in (a) is the FOV of the lower panels.  Bottom row: the 304 {\AA} base-difference images are plotted as background grayscale images and 1600 {\AA} intensity maps as contours.
} 
\end{figure}

\subsection{Flare Ribbons}

In general, ribbons are best identified in 304 {\AA} and 1600 {\AA} images, so we investigate the ribbon dynamics in further detail using these images.
In Figure 3, we plot the 304 {\AA} intensity over the HMI longitudinal magnetograms (top panels) and the 1600 {\AA} intensity over the 304 {\AA} images (bottom). 
These 304 {\AA} images are base-difference images relative to the image at 04:10 UT. Note that times selected for display in the top panels differ from those of the bottom panels. From left to right, the images are at preflare, flare maximum, and postflare times (top panels), and three times on and after the maximum (bottom). 

The circular ribbon is narrow in the southern section (Fig. 3a), and becomes wider with time (Fig. 4bc). The circular ribbon in the northwestern section is very faint in the 304 {\AA} images throughout, and as far as we could trace, that part of the ribbon is inactive. The most noticeable change occurs in the northeastern part of the circular ribbon (NCR). In the initial phase (Fig. 3a) NCR is faint. At flare maximum (Fig. 3b), NCR is broad and has maximally expanded northward, and the inner flare ribbon (FR1) stands out. In the late phase (Fig. 3c), another northeast ribbon appears south of the previous location of NCR. For clarity we tentatively distinguish this flare ribbon (FR2) from the pre-existing circular ribbon (NCR), although previous studies have assumed that circular ribbons are flare ribbons.

The bottom panels of Fig. 3 provide a closer look at the evolution of FR1 and FR2, clarifying their identity as flare ribbons. 
The 304 {\AA} maps initially show the ribbons and later flare loops as well, while the 1600 {\AA} emission keeps appearing at the footpoints of the 304 {\AA} loops. In that sense, we trust the 1600 {\AA} intensity more as a ribbon indicator in the postflare phase. At the start of the flare (Fig. 3d), both NCR and FR2 coexist in the 304 {\AA} and the 1600 {\AA} images. 
During the flare, FR2 approaches NCR, and later NCR disappears and only FR2 is left (Fig. 3e). In the area between NCR and FR2, the 304 {\AA} images show dimming. This means that the closed fan fields rooted between NCR and FR2 open up and are consequently disconnected from the active reconnection site. In effect, FR2 merged with the new NCR. 
With time, the eastern thin portion of FR2 mainly extends eastward (Fig. 3f).  The motion of the conjugate ribbon, FR1, is less clear, but also tends to remain as conjugate footpoints of the 304 {\AA} loops  extending eastward. Therefore, these ribbons evolve like an expanding postflare arcade typically seen in two-ribbon flares.

\begin{figure}[tbh]  
\plotone{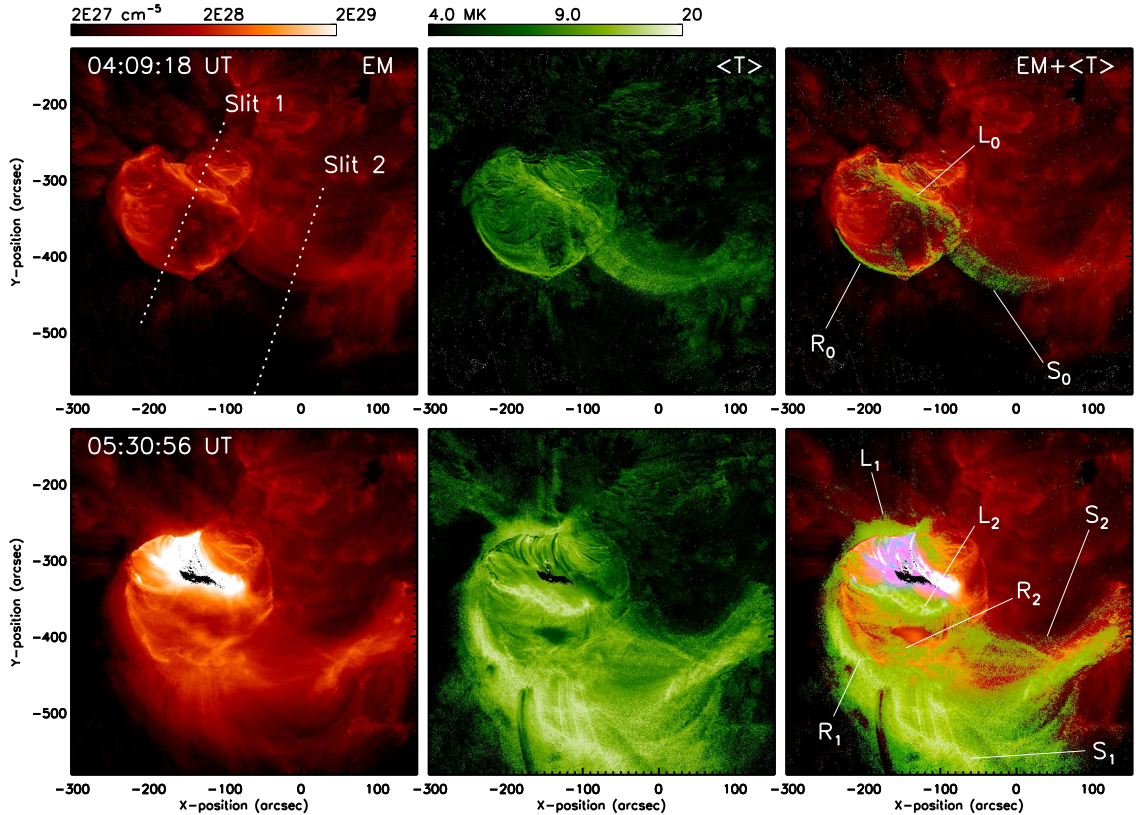}
\caption{Spatial distributions of EMs and temperatures at a preflare (top row) and a postflare (bottom row) time. The left column shows the EM maps, and the middle, the  temperature maps. Composites of EM and temperature are shown in the right column with high temperature ($\geq$50\%) regions (green colortable) plotted over the EM maps (red). Two slit positions and three regions of interest are also denoted: $L$, $R$, and $S$. The accompanying animation shows the temperature and EM distributions determined from the two sets of AIA EUV images with and without AEC, respectively.
}
\end{figure}

\section{DEM Analysis}

We convert the AIA EUV intensities to temperature and emission measure (EM) by performing a DEM inversion analysis. Especially we pay attention to three key regions: flare loops ($L$), the southern part of the circular ribbon ($R$), and the outer spine halo ($S$).
Because this analysis uses six EUV lines representing mostly coronal plasma, therefore the result does not necessarily reproduce the ribbon dynamics that we investigated above using the 304 {\AA} and 1600 {\AA}.
For instance, most EUV images initially show the southern circular ribbon, but later the expanding fan halo may overlie the ribbons along the line of sight and the inversion result may include contribution from the fan plasma. 
The effect of overlapping coronal plasma should be taken into account in interpreting the result for $R$. Although the analysis result for $L$ could suffer the same effect, it is a much stronger source and contribution from the overlying fan halo may be negligible. $S$ is a weak source, but it is relatively isolated from other features that could contribute along the line of sight. 

Another technical issue is that the exposure control may work differently for each region of interest. It is typical to choose the frames with automatic exposure control (AEC) to reduce the problem of the saturated pixels in the flare region ($L$). However, in faint regions without saturation, such as the outer ribbons ($R$) and outer spine ($S$), the EUV frames without AEC are better because they were taken with longer exposure times ($\sim$2 s) and show more details. We therefore decided to use two sets of AIA data, images with AEC for $L$ and images without AEC for $R$ and $S$, at the cost of degrading the time cadence by a factor of about two.  The differences between the DEM results with and without AEC can be checked in the  animation accompanying Figure 4.

We first determine the DEM in each pixel using the procedure by Cheung et al. (2015), and calculate the EM maps by simply adding up the DEM maps over the temperature grid ($5.5 \leq \log T \leq 7.5$). Similarly the average temperature ($\langle T\rangle$) maps were created by adding the temperature grid weighted by DEM, i.e., $\langle T\rangle =\sum_i T_i {\rm DEM}_i/\sum_i {\rm DEM}_i$. However, other types of average temperature will be introduced in \S3.3.  When we study the overall structure in \S3.1 and \S3.2, we use the AIA EUV data with AEC only (Figures 4 and 5). For study of the local thermal properties, we use the combination of the data with and without AEC (Figures 6 and 7).

\subsection{Thermal Structure}

Figure 4 shows the EM maps (left column),  $\langle T\rangle$ maps (middle), and composite maps of EM and $\langle T\rangle$ (right) at a preflare time (top) and a postflare time (bottom).  Both EM and $\langle T\rangle$ maps are in logarithmic scale in the ranges specified in the color bars. In the composite maps, only the regions with $\langle T\rangle$ higher than 9 MK (green) are overplotted in the EM maps (red). Even at preflare times, the circular ribbon is clearly visible, especially in the $\langle T\rangle$ maps, where the circular ribbon (denoted initially as $R_0$) appears as a very narrow lane of enhanced $\langle T\rangle$ lying at the edge of the EM-enhanced region,
and becomes a little thicker during the flare. The outer spine halo ($S_0$) is more clearly seen in the $\langle T\rangle$ maps than in the EM maps but emits less than the fan region, because the halo is filled with lower density plasma.  The loops marked $L_0$ initially connected the inner and outer flare ribbons with the highest $\langle T\rangle$, even during the preflare phase. 
To indicate the expanding hot features with time, we introduce more notations: $L_{1,2}$, $R_{1,2}$, and $S_{1}$, with 1 and 2 denoting both sides of one entity rather than two separate features. 

The small initial flare arcade $L_0$ expands as represented by two temperature-enhanced fronts moving away from the PIL in the northward ($L_1$) and southward ($L_2$) directions and thickening. Enhanced EM appears between $L_1$ and $L_2$, consistent with postflare arcade development. 
As mentioned earlier in this section, both features may be mixed with the fan halo features. 
In $R$ there was initially the southern ribbon only,  $R_0$, which gets wider with time. We mark its expansion by both the outward $R_1$ and inward $R_2$ moving hot fronts. $R_1$ may be associated with the actual expansion of the fan, as demonstrated in other events (Lee et al. 2016ab), and $R_2$ may be due to chromospheric evaporation into newly reconnected field lines. Likewise, as the overlying fan halo expands, $R_2$ may include not only the ribbon component but part of the fan.
Again this expansion of the fan halo is more obvious in the $\langle T\rangle$ maps, and the $\langle T\rangle$-enhanced regions always lie at the edge of the EM-enhanced regions in the direction of expansion.  $S$ is a weak source but has no significant sources around it, so it would suffer less contamination from other features. $S_1$, which marks the outer edge of the spine halo, is significantly south of $S_0$, the preflare spine location.

\begin{figure}[tbh]  
\plotone{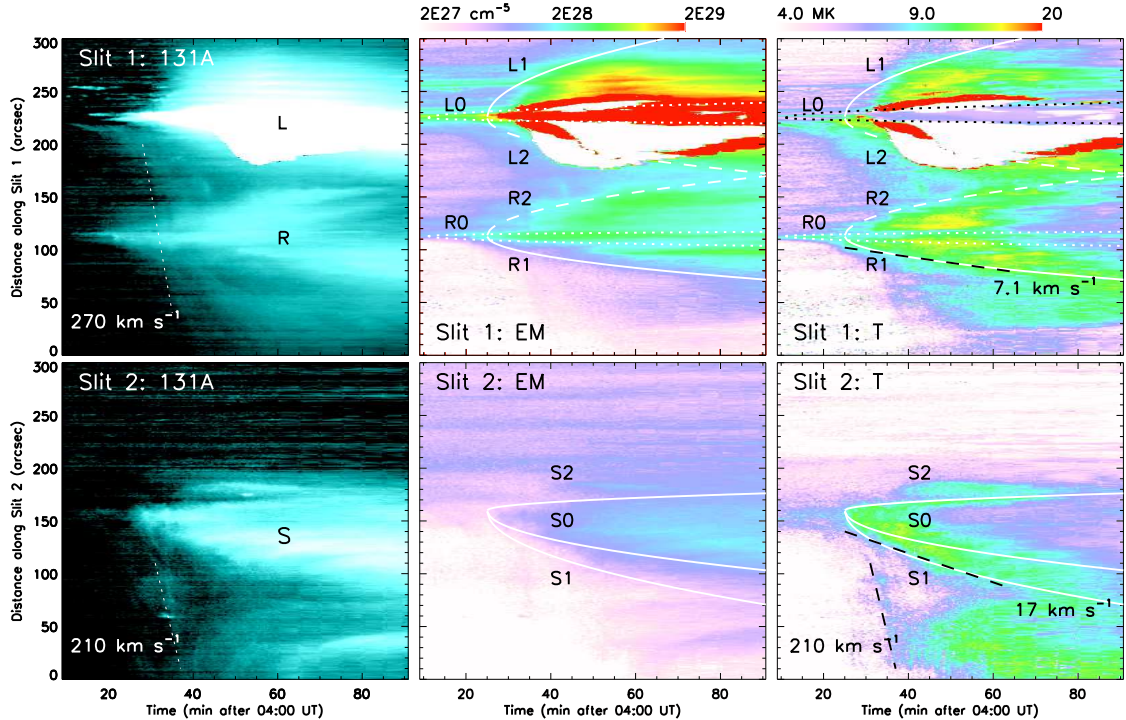}
\caption{Time-distance stackplots of the 131 {\AA} images, EM maps, and $\langle T\rangle$ maps. The plots in the upper (lower) panels are constructed from slit 1 (2). The solid white lines denote the envelopes of the features $L$, $R$, and $S$ defined in Figure 4. Eruption features are denoted with dashed lines together with the estimated speeds.  Here the straight guidelines with speeds denoted nearby represent the linear fits to the expanding features. Other guide lines are the general fits to the envelope of the expanding features.
The white patterns inside $L$ in the second and the third columns are the locations where the DEM inversion failed.
}
\end{figure}

\subsection{Expansion and Eruption}

Figure 5 shows the time-distance ($t$--$d$) stackplots constructed from the two slits denoted in Figure 4.  The $x$-axis is time and the $y$-axis is the distance along the two slits, increasing from the south end to the north end. The left column shows the stackplot of the 131 {\AA} intensity. We again confirm the expansion of the flare arcade (denoted $L$), the circular ribbon ($R$), and the outer spine halo ($S$). Furthermore the expansion of $R$ starts simultaneously with that of $L$, which implies that they are magnetically connected to a common energy release source. The eruption features are much fainter than the expansion features. The first one is detected in slit 1 with a speed of 270 km s$^{-1}$ (dashed line in Fig. 5a). It is unclear when this feature started because of the bright emission in the flare loops. Assuming that it started from the fan surface, we estimate the starting time as $\sim$04:25 UT.  In slit 2, we find another eruption at a speed of 210 km s$^{-1}$ starting at 04:29 UT. Considering the uncertainty in this estimation, both features detected in the two slits could be part of a single coherent eruption, and their starting times between 04:25 UT and 04:29 UT roughly coincide with starting time of expansion features in all regions. Liu et al. (2019) also noticed a transition in the surrounding magnetic field evolution around 04:28 UT, signifying that the eruption had global-scale effects.

The middle and the right panels show $t$-$d$ stackplots of EM and $\langle T\rangle$. Since EM or $\langle T\rangle$ may evolve without any changes in the magnetic structure, these features may not necessarily show the same dynamics of the field lines seen in the EUV images.  Nevertheless, the expanding halo seen in the 131 {\AA} images is also seen in both $T$ and EM maps. Once again, the highest $\langle T\rangle$s are found at the edge of the high-EM region, indicating that energy is newly injected into the moving front. On the other hand, the eruption itself is much less obvious in $\langle T \rangle $ and EM maps. Only the $\langle T \rangle$ plot of slit 2 shows motion roughly perpendicular to the outer spine as fast as 210 km s$^{-1}$, which must represent hot plasma carried away by the erupting magnetic fields.

While $L_2$ keeps expanding southward, $L_1$ ceases to expand by 04:50 UT. It is likely that $L_1$ is bounded by the separatrix, so the northern flare ribbon merges with the circular ribbon there. The relative location and timing of the enhanced temperature in $L$ may yield insight  into the flare loop heating and the associated magnetic reconnection (cf. Wyper et al. 2018).  Unfortunately, EUV intensities in the hottest part of the flare arcade are mostly saturated, and DEM analysis in those pixels failed. Because $R_1$ is coincident with the circular ribbon seen in 304 {\AA} and 1600 {\AA}, it marks a section of the fan footpoints as it expands outward.  The expansion speed of the hot envelope (outlined by $R_1$) is estimated to be $\sim$7.1 km s$^{-1}$. $R_2$ also appears to expand at a similar speed. 
 In view of the slow speed, it may be upward motion through evaporation caused by reconnection, filling the loops with hotter, denser plasma. The outer spine halo also expands at the speed of $\sim$17 km s$^{-1}$ from $S_0$ to $S_1$.

\begin{figure}[tbh]  
\plotone{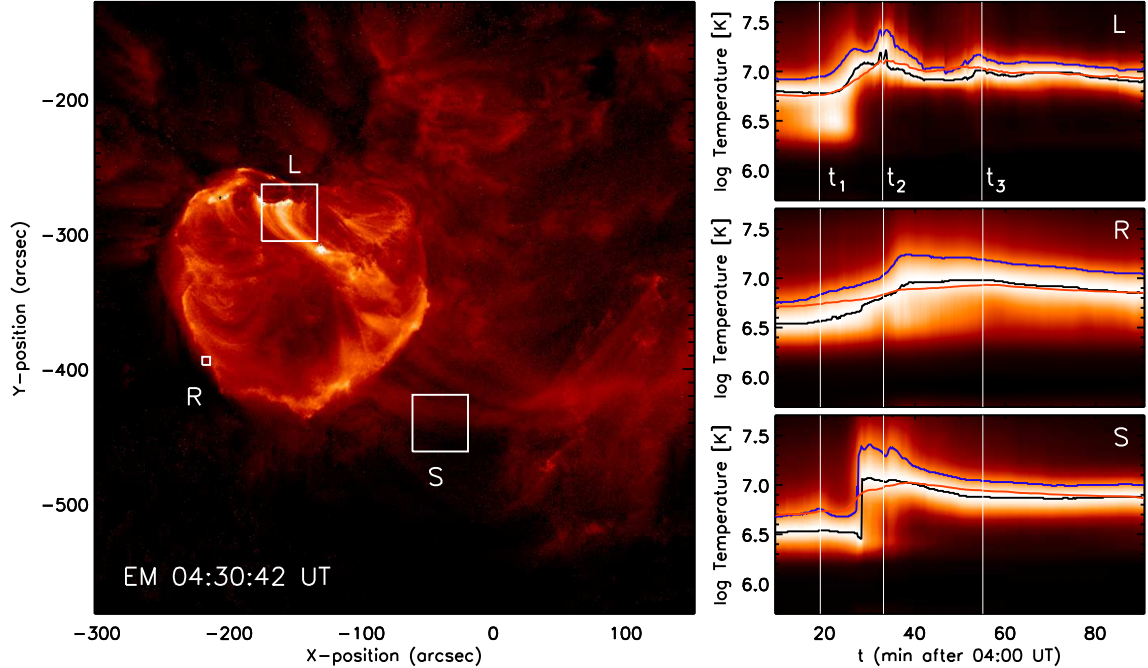}
\caption
{Thermal evolution  of the circular ribbon flare in regions of interest. (a) Three regions of interest --- flare loops (L), circular ribbon (R), and the outer spine (S) --- are marked by white boxes on the EM map at 04:30 UT. (b)-(d): normalized DEMs as functions of time in the three local regions. The curves colored red, blue, and black are three different kinds of average temperatures (see text). 
The three vertical guide lines represent $t_1$--$t_3$ as defined in Figure 1. 
}
\end{figure}

\subsection{Local Thermal Activities}

We determine the local thermal activities using local DEM evolution in the regions of interest denoted in Figure 6(a). As aforementioned, we use two sets of the AIA EUV data depending on location. For the flare region, $L$, we use the frames with AEC to reduce the problem of the saturated pixels, and for the faint regions, $R$ and $S$, the EUV frames without AEC are used to take advantage of the longer exposure time. Note also that we used the pixels within a very small box for the ribbon $R$, so as to exclude non-ribbon regions in calculating the ribbon $\langle T\rangle$.
The three panels labeled (b)--(d) in the right side of Figure 6 show the relative DEMs from $L$, $R$, and $S$ as functions of time ($x$-axis) and the logarithmic temperature ($y$-axis). The DEMs are normalized to the maximum DEM at each time interval so that we can easily check at which temperature DEM has its maximum. The three vertical guidelines labeled $t_1$--$t_3$ defined in Figure 1 are plotted here again for comparison.

The red lines in the right column of Figure 6 are the DEM-weighted average temperatures, $\langle T\rangle$. Although this is the most commonly adopted form of average tempaerature, it may dismiss rapid temperature variations, especially between $t_1$ and $t_2$. A careful examination of the DEM at individual times shows that the plasma is multithermal, so a simple average would obscure impulsive heating events.    
We thus devise two alternative temperatures. One is the temperature at which the DEM has its maximum at each time, plotted as black lines in Fig. 6b-d. This temperature more clearly shows the rapid changes between $t_1$ and $t_2$, so that temperature variations in $R$ and $S$ as impulsive as those in $L$ can be found. This average temperature is insufficient in some cases to represent impulsive plasma heating, because at such times the DEM not only peaks at a higher temperature but also is distributed over a wider temperature range. To represent such behavior, we choose temperatures at which the DEM reaches 75\% of its maximum, and plot them as blue lines in Fig. 6b-d. This approach better captures the impulsive nature of the heating events as well as the higher temperatures that the plasma reaches.

Figure 7 shows the local EMs in panel (a) and the local average temperatures in (b)-(d) as functions of time. 
The local EMs are calculated by adding up the DEMs over the boxed areas, excluding the pixels with no DEM solutions, and dividing the sum by the number of effective pixels, so that the numbers represent average EM per pixel independent of the box size. 
Also shown are the GOES soft X-ray lightcurves, which appear to be similar to the EM curve in $L$. Therefore region $L$ apparently produced the dominant EM. We find that the EM of $L$ changed its slope at the times corresponding to the three peaks, $t_1$--$t_3$, and dominated in the extended late phase. The EMs from $R$ and $S$ also increased, starting at $t_1$ and rising most impulsively toward $t_2$, but both were much smaller than the EM in $L$ and contribute little toward the total EM.

\begin{figure}[tbh]  
\plotone{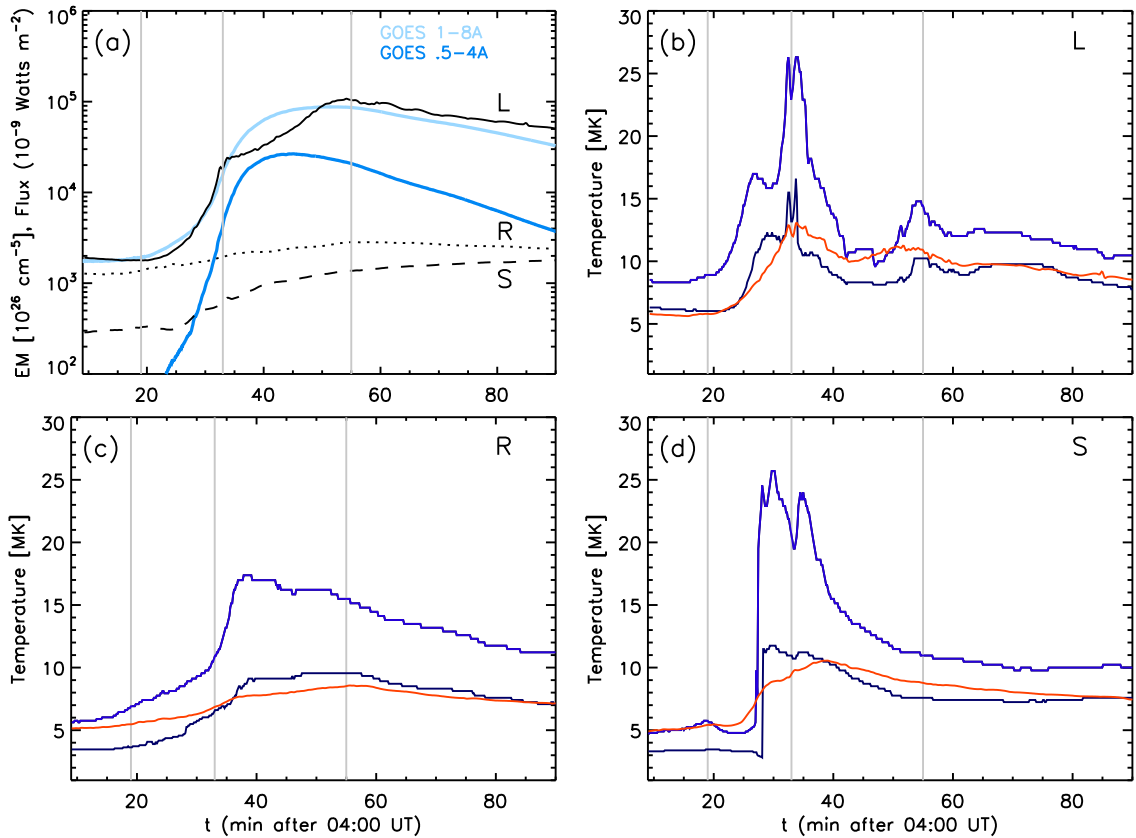}
\caption
{Local thermal evolution. (a) Local EMs of the three regions, $L$, $R$, and $S$ together with the GOES soft X-ray lightcurves.  (b)-(d) Local average temperatures of the three regions as functions of time. The red, blue, and black lines denote average temperatures $\langle T\rangle$ as defined in the text.
}
\end{figure}

The three kinds of $\langle T\rangle$ are plotted in Figure 7(b)-(d) as functions of time for the three local regions, $L$, $R$, and $S$. The DEM-weighted temperature (red lines) in $L$ starts to rise at $t_1$, reaches the main peak at $t_2$, and exhibits another peak at $t_3$, whereas both those in $S$ and $R$ reach their maxima around $t_2$ and monotonically decrease afterwards. Specifically, the temperature in $L$ shows more than two peaks: the first one between $t_1$ and $t_2$, the maximum at $t_2$, and the third at $t_3$. Interestingly, the temperature in $S$ also rises very rapidly, reaching a maximum at the time of the first peak in $L$, which also coincides with the start of the GOES X-ray 1--8 {\AA} lightcurve. 

The DEM-weighted temperature profile in $R$ significantly differs from the profiles in $L$ and $S$, in that it slowly reaches a single peak around $t_3$ and decays slowly thereafter. This is probably due to the fact that the plasma in $R$ is of chromospheric origin, while the others are coronal.
The most noteworthy result is that $t_3$, defined based on the EUV fluxes (Fig. 1), coincides with the onset of the secondary temperature rise in $L$, representing additional thermal heating in the decay phase that is absent in either $R$ or $S$. Therefore we conclude that flare reconnection and the associated heating continued and became stronger for a while. This result is consistent with the Hock et al. (2019) hypothesis for late-phase emission, but conflicts with the idea that the extended EUV late phase must be due to heating in large loops outside the fan (Sun et al. 2013).

\section{Discussion and Conclusion}

\begin{figure}[tbh]  
\plotone{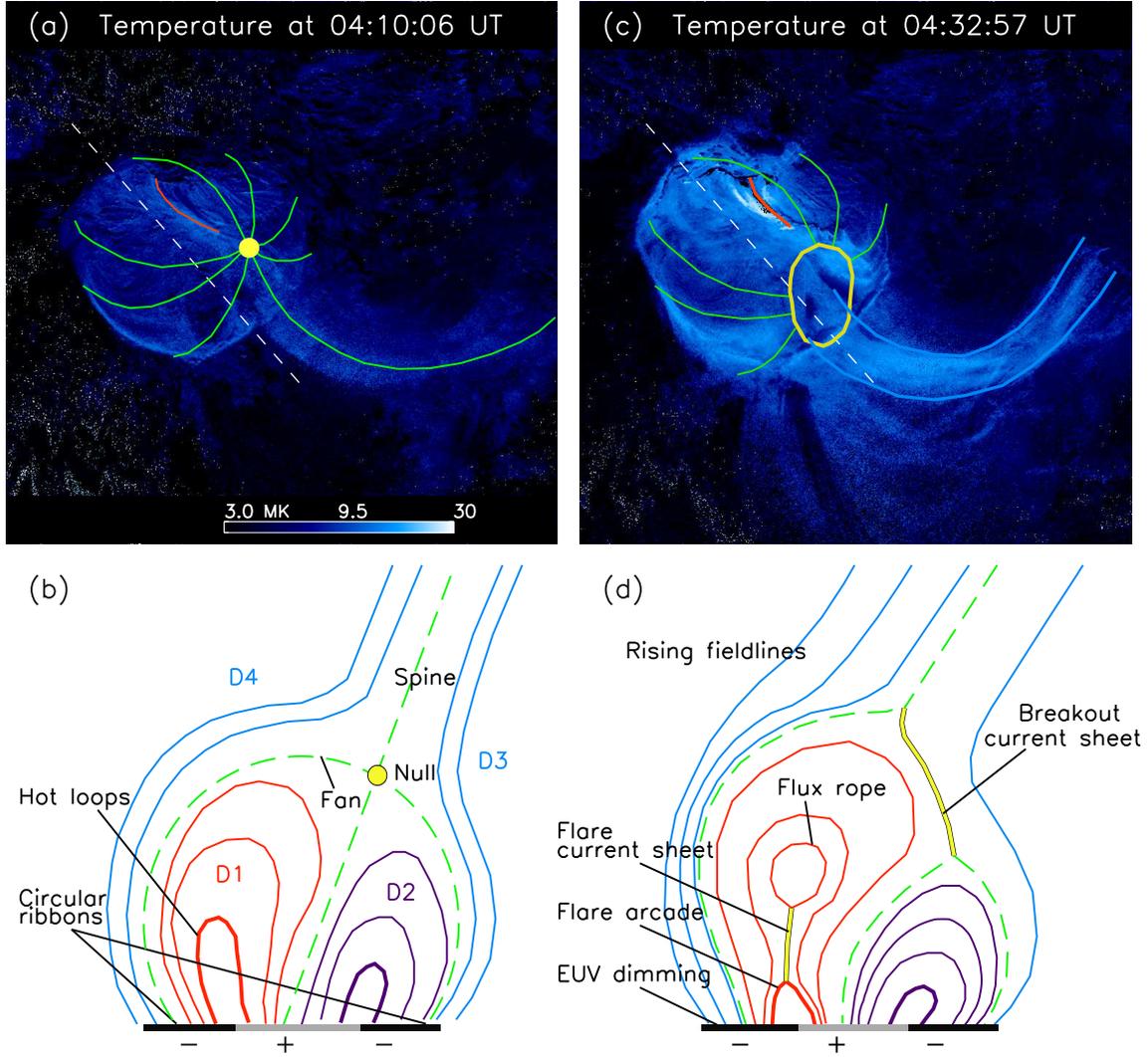}
\caption{Illustration of the proposed magnetic structural change from the preflare phase (a, b) to the pre-eruption phase (c, d). In (a) and (c), the background images are average temperature maps with the scale denoted in the bottom of (a). The overplotted lines are schematic representations of the fan and outer spine field lines (green), the flare loops (red), and the null (yellow dot), which evolves into a breakout current sheet (yellow ellipse). (b) and (d) show 2D drawings of the magnetic structure in a vertical plane oriented in the northeast-southwest direction (white dashed line), using the same color convention as in (a, c) and including our interpretation of the key topological elements: fan, spine, current sheets, flare arcade, EUV dimming, and circular ribbons.  For more details, see text.
}
\end{figure}

We have studied the spatio-temporal evolution of a circular ribbon flare, the SOL2014-12-17T04:51 in NOAA AR 12242, with emphasis on the EUV diagnostics for $\langle T\rangle$  and EM. The picture of the circular ribbon flare emerging from this analysis is that the fan-spine structure is activated first, after which the system is heated and reconfigured until the eruption.  The key point is that a stressed null point always transforms itself to a current sheet, so that the confined flux under the fan structure more easily opens up through interchange reconnection with the external field. Removing the overlying flux allows a flux rope formed deep within the fan to erupt (see also Panesar et al. 2016), while the interchange reconnection itself produces narrow outflows (jets). This scenario is based on the breakout model outlined in \S1, in which  the current sheet at the distorted null is denoted the breakout current sheet. We have not directly detected this breakout current sheet in any images presented here, but such a thin patch is highly unlikely to be visible when viewed on the disk (however, see Kumar et al. 2018, 2019 for examples of breakout current sheets seen in limb events). 

This fundamental behavior is illustrated schematically by Figure 8. Top panels show temperature images at selected times during the preflare (a) and pre-eruption (c) stages. The overplotted lines represent fan and outer spine field lines (green) and hot/flare loops (red). The 2D cartoons in (b) and (d) visualize slices of the magnetic structures in (a) and (c) in a vertical plane oriented in the northeast-southwest direction (along the white dashed line), and include our interpretation of the key topological elements based on the breakout model. For illustration, we show four domains, D1--D4. D1 and D2 are separated by the spine (green dashed line in (b) and (d)) in these 2D cartoons, but actually comprise one connectivity domain containing closed fields only. Likewise, D3 and D4 jointly comprise a single connectivity domain filled with open field lines. D1--D2 and D3--D4 are separated from each other by the fan surface (i.e., separatrix, marked by green dashed lines). The cartoon in (b) shows a fan/spine structure with a null (yellow dot) with more flux in D1 than in D2, so that the spine is tilted toward the south, resembling the potential field extrapolation (see Figure 2 of Liu et al. 2019). 
As the filament channel at the core of D1 (inside the hot loops) accumulates more shear, the sheared flux expands, pushing the overlying flux outward and deforming the null point into a breakout current sheet (upper yellow line), as shown in (d). This brings D1 flux into contact with the opposite polarity flux in D3, initiating interchange reconnection and removing flux from both domains.  This enables the D1 flux to continue rising, forming the flare current sheet (lower yellow line) beneath and creating a growing flux rope and the flare arcade through flare reconnection.  The expanding flare arcade in D1 and the rising open fields in D3 are more apparent in the temperature images than in any of the EUV images (Figures 4-5). The rapid loss of material along the newly opened field lines manifested by the EUV dimming (Figure 3) is also indicated in (d). Under the scenario depicted here, we explain most of the observed activities as follows.

\begin{enumerate}

\item 
In the preflare phase, the circular ribbon was lit up prior to other activities. In a fan-spine structure like the current AR, the most plausible mechanism for this activation is breakout reconnection, after the null point is transformed into a breakout current sheet. Weak outflows and the observed displacements of hot loops above the fan support this scenario. Slow interchange reconnection between the closed fan loops and the effectively open external field produced mild heating, particle precipitation to the fan footpoints  (generating the circular ribbon), and weak outflows starting at $\sim$04:20 UT. Therefore slow breakout current sheet reconnection started well before the flare, consistent with the recent jet simulations and observations (Wyper et al. 2018, Kumar et al. 2019). 

\item 
The fan halo expansion as detected in the $t$-$d$ stackplots of $\langle T\rangle$ and EM may have started when sufficient breakout reconnection altered part of the northern separatrix location. The slow breakout reconnection produced some energetic electrons and heated the plasma on newly reconnected flux tubes, widening the visible area occupied by the circular ribbon. The higher $\langle T\rangle$ layer appeared at the edge of the expanding EM ribbons, suggesting that more heating and/or particle acceleration and thus more evaporation occurred in the newly reconnected flux tubes.

\item
Although we did not directly detect the flux rope formed under the dome, the flare arcade signifies that a flux rope formed above the filament channel around the PIL. Continued flare reconnection built up the flux rope, causing it to rise and ultimately interact with the overlying fan field, thus initiating the main impulsive phase of the flare. The outflows were also hardly visible, because this event occurred near disk center and any ejecta would be directed along the line of sight. The eruptions were only detected in the $t$-$d$ stackplot of 131 {\AA}, while they are not so clear in the EUV images. Most of the flare energy is released when the flux rope interacts with the overlying fan, enabling eruptive outflows to escape, probably leading to the CME described in \S2.

\item
The location of the outer flare ribbon (FR2) inside the northern circular ribbon (NCR) confirms that a flux rope formed along the filament channel inside the fan surface. Because the erupting portion of the PIL is situated close to the northern boundary of the fan, FR2 can't travel far from its initial position. 
Expansion of the flare ribbons parallel to the PIL during the impulsive phase also reflects the spatial extension of fast flare reconnection in the overlying flare current sheet.  The dimming region appearing between FR2 and NCR must correspond to the footpoint of the erupting flux rope. 
 
\item
A typical flare arcade appears after the flare ribbons (FR1 and FR2) brighten substantially, indicating that fast flare and breakout reconnection have already begun. 
Since the impulsive phase of the flare occurred as the flux rope reconnected with the external field through the breakout current sheet, the flux rope could partially or fully escape as an eruption. 

\item
Identifying the location of the additional late-phase heating inside the fan surface is a new result consistent with the explanation offered by Hock et al. (2019), but inconsistent with the suggestion that large loops outside the fan surface are responsible for the extended phase (Sun et al. 2013, Masson et al. 2017).

\end{enumerate}

In this interpretation, it becomes clear how the eruption initiated and ended successfully to produce a CME in this event. 
The null point--to--breakout current sheet transformation is the essential process in the whole event. 
The energy stored beneath the null point must be the main driver for the eruptive circular-ribbon flare, and the null point reconnection is not necessarily the dominant process. A circular ribbon flare may proceed in an eruptive manner despite its initial confined structure because a breakout current sheet can form at the null point, thus allowing a jet or a flux rope to escape via breakout reconnection.

\acknowledgments
This work was supported by the NASA grants, 80NSSC18K1705, 80NSSC17K0016, and 80NSSC18K0673, and NSF grants, AGS 1821294 and AGS 1927578. JTK thanks S. Antiochos for illuminating discussions.

\end{document}